# CRITICAL FLUCTUATIONS OF A CONFINED BINARY MIXTURE

*Małgorzta Zubaszewska[1], Andrej Gendiar[2], Tomasz Masłowski[1]*

[1]*Institute of Physics, University of Zielona Góra, ul. Prof. Z. Szafrana 4a,
65-516 Zielona Góra, Poland*
[2]*Institute of Physics, Slovak Academy of Sciences,
SK-845 11 Bratislava, Slovakia*
[1]*mzubaszewska@o2.pl,*  [2]*andrej.gendiar@savba.sk,*  [1]*T.Maslowski@if.uz.zgora.pl*

**Abstract.** Exploiting the mapping between a binary mixture and the Ising model we have analyzed the critical fluctuations by means of the density-matrix renormalization group technique. The calculations have been carried out for a two-dimensional Ising strip subject to equally strong surface fields. It was found that the critical Casimir force has significantly different behavior on opposite sides of the capillary condensation line, especially below the critical temperature. It can be concluded that in real binary mixtures the most attractive force appears at temperatures near $T_c$ and at reservoir compositions slightly away from the critical composition.

## 1. Introduction

The confinement of long-ranged critical fluctuations in the vicinity of second-order phase transition in a binary mixture generates effective forces arising between confining surfaces, known in the literature as critical Casimir forces [1-4]. They acquire universal features upon approaching a critical point of the medium and become long ranged at criticality. Moreover, such forces can also act among particles immersed in a critical mixture [5]. This is realized in binary liquid mixtures close to their critical point $T_c$ which belong to the universality class of the Ising model [6-9].

Fig. 1 presents the phase diagram of a binary mixture representing two-component liquid where the components are denoted by A and B. There is one mixed state above the critical temperature $T_c$ and two coexisting phases below $T_c$ which are separated into A-rich phase and B-rich phase. The metastable states have not been presented in Fig. 1. The deviation of the difference of the chemical potentials of the two components of the mixture from the value at criticality corresponds to the bulk magnetic field of the Ising model. The surface field corresponds to the overall preference of a surface for one of the two phases (component A rich or component B rich).

Another surface effect, becoming increasingly important when the size of system is decreased, is the capillary condensation. Its essence lies in the shift of the phase coexistence line to a finite value of the bulk magnetic field $H_0(T; L)$ as the combined effect of identical boundary fields and confinement. For short-range boundary fields this line scales for large $L$, according to the Kelvin equation as $H_0(T; L) \sim 1/L$ [10]. This phenomenon is analogous to the capillary condensation for one-component fluid confined between parallel surfaces, where the gas-liquid transition occurs at a lower pressure than in the bulk.

In the present paper the critical Casimir force as a function of temperature and the bulk field is studied for the two-dimensional Ising model. We assume that boundaries of the strip are equal and belong to the so-called normal transition surface universality class $(+, +)$ [11], which is characterized by a strong effective surface field acting on the corresponding order parameter of a system. Normal transition surface universality class is an appropriate characterization of a critical binary mixture in the presence of an external wall. For these systems the critical Casimir force is expected to be attractive for all thermodynamic states [1-5].

Our goal is to determine the dependence of critical Casimir force on the bulk field and temperature. We are interested in such a range of parameters where the properties of the confined fluid near bulk criticality are particularly rich since the combined effects of finite-size and specific wall-fluid interactions. Hence, we would like to assign a range of thermodynamic parameters for which the force is the most attractive.

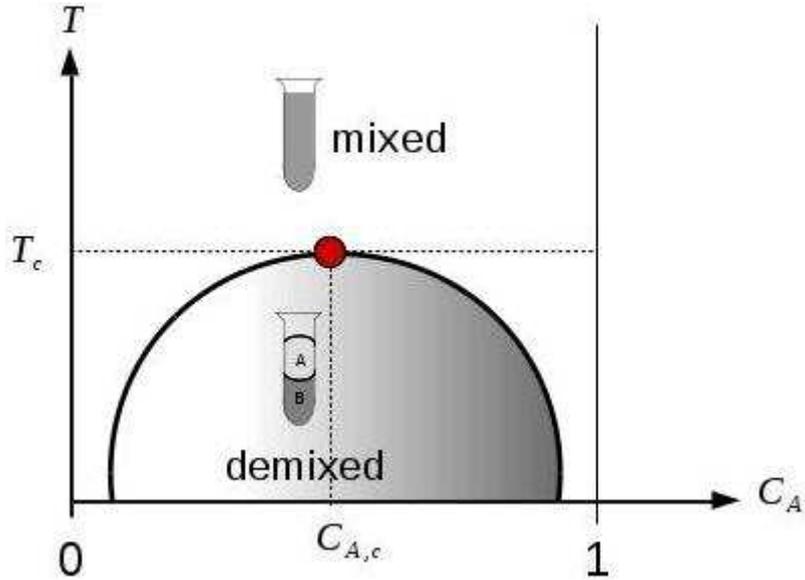

Fig. 1. Schematic phase diagram of a binary mixture of liquids A and B with an upper critical point represented by the dot mark. $C_A$ denotes the concentration of the liquid A while $C_{A,c}$ is its critical concentration. According to our results, the darker the interior shading, the more attractive the Casimir force becomes. The concentration of the liquid B is equal to $1 - C_A$.

## 2. Model

We consider a two-dimensional Ising model on the square lattice in a slit geometry $M \times L$ subject to identical boundary fields. The energy for a configuration $\{s\}$ of spins is given by the Hamiltonian:

$$\mathcal{H} = -J \left( \sum_{\langle ij, i'j' \rangle} s_{i,j} s_{i',j'} - h_1 \sum_{\substack{surface \\ spins}} s_{i,j} - h \sum_{\substack{all \\ spins}} s_{i,j} \right), \quad (1)$$

with $J > 0$ and $s_{i,j} = \pm 1$, where $(i,j)$ labels the site of the lattice. The first sum is over nearest neighbors, while the second sum is performed over spins at the both surfaces. We are interested in the limit $M \to \infty$ with finite $L$. Bulk and surface fields are measured in the units of the coupling constant J and the distances are measured in units of the lattice constant.

## 3. Method

In order to calculate the free energy of a system, we use the density-matrix renormalization-group method (DMRG) originally introduced to study ground-state properties of quantum-spin chains [12-14]. In spite of the name, the method has only some analogies with the traditional renormalization group being essentially the numerical, iterative basis, truncation method.

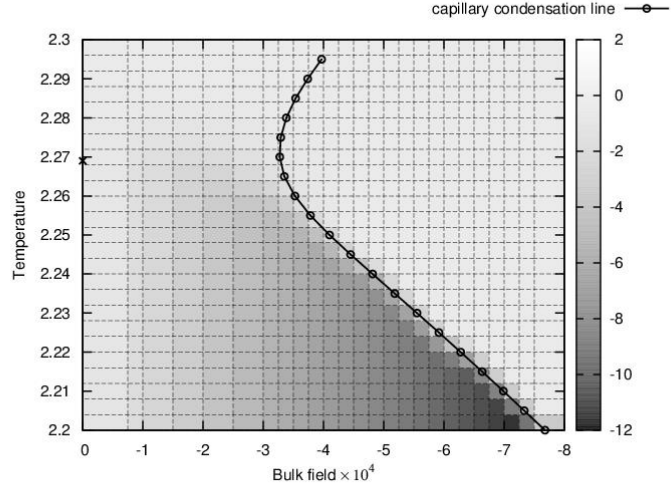

Fig. 2. The critical Casimir force as a function of the bulk field and temperature. The solid line represents the capillary condensation line $H_0(T; L = 301)$. The thick cross symbol represents the bulk criticality at $h = 0$ and $T = T_C \approx 2.269 \frac{J}{k_B}$.

Next the DMRG was adapted by Nishino for two-dimensional classical systems at nonzero temperatures [15-19]. The total free energy $f$ per site is obtained from the largest eigenvalue $\lambda_0$ of the effective transfer matrix

$$f(L, T, h, h_1) = -\frac{k_B T}{L} \ln \lambda_0. \tag{2}$$

The linear dimension of the total effective transfer matrix is $4m^2$. The $m$ parameter determines the dimension of the subspace of system states, which are kept during subsequent steps the higher $m$, the greater the precision of results. In the present case we have found that the value $m = 50$ is sufficient to guarantee very high accuracy. Apart from the direct neighborhood of the critical point our results for the free energy hold 12 significant digits of accuracy.

In order to obtain the critical Casimir force we calculate first the excess free energy per unit length in the (1,0) direction $f_{ex}(L) \equiv [f(L) - f_b]L$, where $f_b = f(L \to \infty, T, h, h_1 = 0)$ is the bulk free energy per spin [10]. For the vanishing bulk field, $f_b$ is known from the exact solution of the two-dimensional Ising model [20]. For nonvanishing bulk fields, in order to get the bulk free energy, the calculations have been performed for the free boundary condition ($h_1 = 0$) when $L$ was increasing up to convergence of the results, which, in practice, meant no more than 2000 lattice constants.

The expression for the critical Casimir force can be defined as [1]:

$$F_C = -\left(\frac{\partial f_{ex}}{\partial L}\right)_{T,h,h_1}. \qquad (3)$$

In turn, the capillary condensation line $H_0(T; L = const)$ can be determined by the maxima of the total free energy [15-19].

## 4. Results

Fig. 2 presents the critical Casimir force for the strip of the width $L = 301$ and the surface fields $h_1 = 8.15$ providing the very strong field limit. The capillary condensation (coexistence) line has a positive slope and is located at $h < 0$ (because the surface field is a positive number).

As one can see, at capillary condensation, the critical Casimir force exhibits a jump from a large value for thermodynamic states corresponding to the demixing phase to a vanishingly small value for those corresponding to the mixed phase. The darker area in the picture, the more attractive the Casimir force becomes. Further details of the critical Casimir force as a function of two arguments can be found in a perspective plot (Fig. 3).

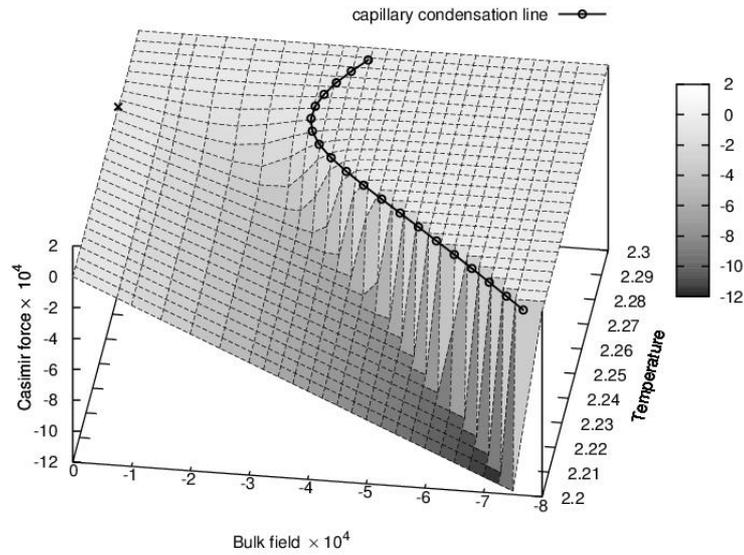

Fig. 3. Three-dimensional plot of the critical Casimir force as a function of the bulk field and temperature.

## 5. Conclusions

The critical Casimir force arises in the thermodynamic description of confined binary fluid mixtures as an excess pressure over the bulk value and is conjugate to the distance $L$ between the confining surfaces. Such a strongly attractive force between two large colloidal particles immersed in a near-critical fluid can have ramifications for aggregation or flocculation of the particles.

Exploiting the mapping between binary mixtures and the Ising model we have applied the DMRG method providing essentially exact numerical results for strips with widths up to $L = 2000$ lattice constants with various boundary conditions.

The major difference in the behavior of the critical Casimir force on both sides of the capillary condensation line appears below the critical temperature. Therefore, our results imply that in real binary mixtures the critical Casimir force at temperatures somewhat below $T_c$ and at reservoir compositions slightly away from the critical composition should be much more attractive than at the bulk criticality.

## 6. Acknowledgment

M.Z. is a scholar within Sub-measure 8.2.2 Regional Innovation Strategies, Measure 8.2 Transfer of knowledge, Priority VIII Regional human resources for the economy Human Capital Operational Programme co-financed by European Social Fund and state budget. This work was supported by COQI APVV-0646-10 and VEGA-2/0074/12.